\documentclass[letterpaper,twocolumn,10pt]{article}
\usepackage{usenix,epsfig,endnotes}
\usepackage{xcolor}
\usepackage{soul}
\usepackage{hyphenat}
\usepackage[hyphens]{url}
\usepackage{hyperref}
\usepackage{graphicx}
\usepackage{booktabs}
\usepackage[normalem]{ulem}
\usepackage{listings}
\usepackage{lstlinebgrd}
\usepackage{amssymb}
\usepackage{multirow}
\usepackage{breakurl}

\newcommand{\scratch}[1]{}
\usepackage{colortbl}
\usepackage{authblk}
\usepackage{algorithm}
\usepackage[noend]{algpseudocode}
\algnewcommand{\LineComment}[1]{\State \(\triangleright\) #1}
\definecolor{gray}{RGB}{181,181,181}
\definecolor{green}{RGB}{0,100,0}
\newcommand{\codehilir}[1]{\textcolor{red}{#1}}
\newcommand{\codehilig}[1]{\textcolor{green}{#1}}

\lstdefinestyle{minipage}
{
  showstringspaces=false,
  frame=single,
  language=C,
  numbers=left,
  basicstyle=\scriptsize\ttfamily,
  xleftmargin=0em,
  framexleftmargin=0em,
  framexrightmargin=-1em,
  escapeinside={(*@}{@*)},
}

\lstdefinestyle{standalone}
{
  showstringspaces=false,
  frame=single,
  language=C,
  numbers=left,
  basicstyle=\scriptsize\ttfamily,
  xleftmargin=2em,
  framexleftmargin=1.5em,
  escapeinside={(*@}{@*)},
}

\makeatletter
\def\lst@makecaption{%
  \def\@captype{table}%
  \@makecaption
}
\makeatother

\usepackage{etoolbox}\AtBeginEnvironment{algorithmic}{\scriptsize}

\begin{document}

\title{Static Exploration of Taint-Style Vulnerabilities Found by Fuzzing}

\author[1]{Bhargava Shastry}
\author[2]{Federico Maggi}
\author[3]{Fabian Yamaguchi}
\author[3]{Konrad Rieck}
\author[1]{Jean-Pierre Seifert}
\affil[1]{Technische Universit{\"a}t Berlin}
\affil[2]{Trend Micro Inc.}
\affil[3]{Technische Universit{\"a}t Braunschweig}

\renewcommand\Authands{ and }

\maketitle


\begin{abstract}
Taint-style vulnerabilities comprise a majority of fuzzer discovered program faults.
These vulnerabilities usually manifest as memory access violations caused by tainted program input.
Although fuzzers have helped uncover a majority of taint-style vulnerabilities in software to date, they are limited by (i) extent of test coverage; and (ii) the availability of fuzzable test cases.
Therefore, fuzzing alone cannot provide a high assurance that all taint-style vulnerabilities have been uncovered.

In this paper, we use static template matching to find recurrences of fuzzer-discovered vulnerabilities.
To compensate for the inherent incompleteness of template matching, we implement a simple yet effective match-ranking algorithm that uses test coverage data to focus attention on those matches that comprise untested code.
We prototype our approach using the Clang/LLVM compiler toolchain and use it in conjunction with {\it afl-fuzz}, a modern coverage-guided fuzzer.
Using a case study carried out on the Open vSwitch codebase, we show that our prototype uncovers corner cases in modules that lack a fuzzable test harness.
Our work demonstrates that static analysis can effectively complement fuzz testing, and is a useful addition to the security assessment tool-set.
Furthermore, our techniques hold promise for increasing the effectiveness of program analysis and testing, and serve as a building block for a hybrid vulnerability discovery framework.
\end{abstract}

\section{Introduction}
\label{sec:intro}
Software exploitation is asymmetric, requiring only a single flaw to compromise a system, and at times a network of systems.
The complexity inherent to contemporary software increases their attack surface, and plays into the hands of attackers.
Consequently, it is imperative that each and every software module be well analyzed and tested, before a release.
This is especially true for applications that routinely handle untrusted user input such as network and data parsers.
Fuzz testing has been the tool of choice for conducting security assessments of these classes of applications.

Although fuzz testing is effective at uncovering software vulnerabilities, it has two practical limitations.
First, fuzzing may encounter coverage bottlenecks such as cryptographic code, and non-atomic comparison operations that limit the test coverage achieved, and impede the discovery of latent vulnerabilities.
Second, several code bases do not contain test harnesses for security-critical program APIs, making thorough testing dependent on writing new test cases.
Writing test cases is a manual process that requires domain-specific knowledge pertaining to the software under analysis.
{\it Even} well-written unit tests do not necessarily permit a thorough systems evaluation.
For example, networking stacks contain asynchronous, and stateful API calls that are invoked in an event-driven fashion.
Without a practical set-up that injects the right sequence of messages, it becomes difficult to test these APIs.
Having said that, simple pre-existing test cases can provide a starting point for a wider exploration of the codebase.

\begin{figure*}[t]
  \centering
  \includegraphics[scale=.6]{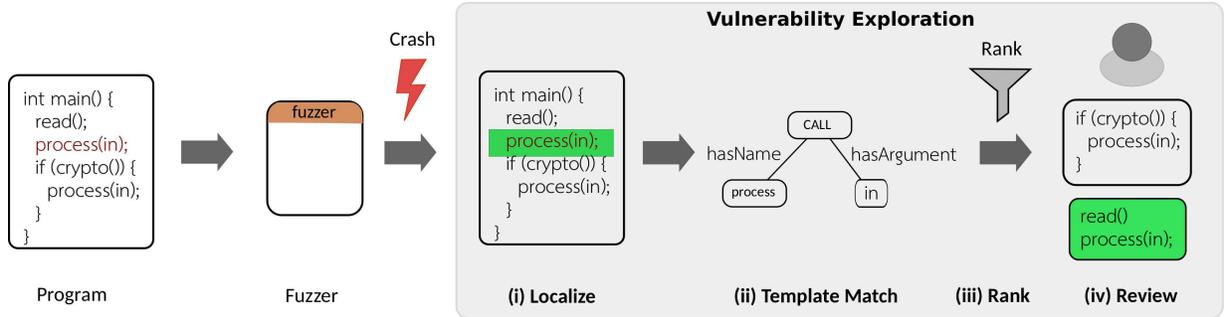}
  \caption{Work-flow of static vulnerability exploration. Templates generated from fault localized code are used to find recurring instances of a fuzzer-discovered vulnerability. The resulting matches are ranked to focus attention on potential recurring vulnerabilities in untested code.}
  \label{fig:workflow}
\end{figure*}

In this paper, we build on the idea that static analysis can perform a broader search for vulnerable code patterns, starting from a handful of fuzzer-discovered program failures.
Our working hypothesis is that {\it any} readily available fuzzable test harness can be used to bootstrap our analysis, reducing the burden of test writing.
Therefore, we begin by fuzzing an existing test harness packaged with a codebase and expect to find a handful of program crashes.
Subsequently, our analysis proceeds in three steps.
First, we narrow down the root cause of the uncovered crashes using a memory error detector such as AddressSanitizer, falling back to execution slice based fault localization when the fault is not memory-based.
Fault localization not only narrows the search for vulnerable code patterns, but also provides syntactic and semantic information about the underlying fault.
Second, we automatically generate vulnerability templates using localized faulty code.
Vulnerability templates encode both syntactic, as well as semantic features of the faulty code, making our approach superior to na{\"i}ve text-based pattern matchers such as {\tt grep}.
Third, we rank matching code snippets (returned by template matching) by using fuzzer test coverage data: Matches comprising untested code is ranked higher than those that do not.
Such a ranking system helps prioritize manual audit of untested code over code that has already undergone fuzz testing.
We use {\tt afl-fuzz}, a contemporary coverage-guided fuzzer, and Clang/LLVM instrumentation and static analysis framework for prototyping our approach.

We evaluate our prototype using a case study of Open vSwitch (OvS), an open-source virtual switch implementation used in data centers.
We chose Open vSwitch because (i) it routinely handles untrusted input, and (ii) its existing fuzzable test cases achieve a test coverage of less than 5\% providing a low assurance on software security.
Results from our case study are promising.
Our prototype has uncovered a potential recurring vulnerability in a portion of OvS that lacked a test harness.
Moreover, in one instance, template matching has proved to be helpful in flagging a recurring vulnerability that originated in an older release of OvS.
This shows that static analysis can not only complement fuzzing, but enable security assessments to be made during software development.
To facilitate independent evaluation, we have open-sourced our prototype, that is available at \url{https://www.github.com/test-pipeline}.

\noindent
{\bf Contributions:}
\begin{itemize}
\item We present an approach to improve the effectiveness of source code security audit that benefits from both the precise diagnostics of a fuzzer, and the breadth of analysis of a static analyzer.
\item We prototype our approach using {\tt afl-fuzz}, a contemporary coverage-guided fuzzer, and the Clang/LLVM compiler toolchain.
Our prototype automatically generates vulnerability templates from a fuzzer corpus, ranking the matches returned by template matching based on novelty.
\item We evaluate our prototype using a case study of Open vSwitch codebase.
Our approached has (i) helped discover one potential vulnerability in a portion of Open vSwitch that lacked a test harness; (ii) facilitated vulnerability checks at an early stage; and (iii) reduced false alarms by 50-100\% in most cases demonstrating that coverage-based match ranking is effective in combating false positives.
\end{itemize}

\begin{figure*}[t]
  \centering
\begin{minipage}[t]{.43\linewidth}
\begin{lstlisting}[style=minipage,framexrightmargin=0em,label=lst:example,caption=A representative fuzzer test harness in which two synthetic denial of service vulnerabilities have been introduced by calling the {\tt abort()} function. Fault localized code is shown in red.]
#include <string.h>
#include <crypt.h>
#include <stdlib.h>
#include <unistd.h>
#define CUSTOM() abort()
void fuzzable(const char *input) {
    // Fuzzer finds this bug
    if (!strcmp(input, "doom"))
(*@ \codehilir{\hspace{4em}abort();} @*)
}

void cov_bottleneck(const char *input) {
    char *hash = crypt(input, "salt");

    // Fuzzer is unlikely to find this bug
    if (!strcmp(hash, "hash_val"))
       CUSTOM(); // grep misses this
}

// Fuzzer test harness
// INPUT: stdin
int main() {
    char buf[256];
    memset(buf, 0, 256);
    read(0, buf, 255);
    fuzzable(buf);
    cov_bottleneck(buf);
    return 0;
}
\end{lstlisting}
\end{minipage}
\quad
\begin{minipage}[t]{.46\linewidth}
\begin{lstlisting}[style=minipage,xleftmargin=2em, numbers=none,label=lst:query, caption=Template derived from fault localized code (green text) is matched against the test harness source code. Match \#2 lists the line of code containing a similar fault pattern that is likely untested by a fuzzer.]
Template matching:
clang-query> match
(*@ \codehilig{  declRefExpr(} @*)
(*@ \codehilig{\hspace{5em}	      to(}      @*)
(*@ \codehilig{\hspace{6em}        functionDecl(}  @*)
(*@ \codehilig{\hspace{12em}       hasName("abort")} @*)
(*@ \codehilig{\hspace{6em}	 )} @*)
(*@ \codehilig{\hspace{5em})} @*)
(*@ \codehilig{).bind("crash")} @*)

Match #1:

test.c:9:6: note: "crash" binds here
            abort();    
            ^~~~~
test.c:8:5: note: "root" binds here
    if (!strcmp(input, "doom"))
    ^~~~~~~~~~~~~~~~~~~~~~~~~~~

Match #2:

(*@ \codehilig{test.c:17:6: note: "crash" binds here} @*)
(*@ \codehilig{\hspace{6.3em}            abort();} @*)
            ^~~~~
test.c:16:5: note: "root" binds here
    if (!strcmp(hash, "hash_val"))
    ^~~~~~~~~~~~~~~~~~~~~~~~~~~~~~
2 matches.
\end{lstlisting}    
\end{minipage}
\end{figure*}

\section{Static Exploration of Vulnerabilities}
\label{sec:method}
Contemporary fuzzers and dynamic memory analysis tools have greatly advanced vulnerability detection and re-mediation, owing to their ease-of-use and public availability.
Since fuzzers and memory analyzers are invoked at runtime, they require a test harness that accepts user input (usually read from a file or standard input), and invokes program APIs against this input.
Therefore, the effectiveness of fuzzing and dynamic memory analysis depends on the availability of test cases that exercise a wide array of program APIs.

In practice, code bases contain test harnesses for only a limited number of program APIs.
This means that, even if fuzzing were to achieve 100\% test coverage (either line or edge coverage) for the set of {\it existing} test harnesses, it does not lead to 100\% {\it program API} coverage.
Furthermore, for networking software, local test harnesses do not suffice, requiring elaborate setups involving multiple software components.

Our work seeks to counter practical limitations of fuzz testing using a complementary approach.
It builds on the idea that the reciprocal nature of static analysis and fuzzing may be leveraged to increase the effectiveness of source-code security audits.
Our key insight is that vulnerabilities discovered using a fuzzer can be localized to a small portion of application code from which vulnerability templates may be derived.
These templates may then be used to find recurring vulnerabilities that may have been either missed by the fuzzer, or are present in code portions that lack a fuzzable test harness.

\paragraph{Motivating Example}
We motivate our research using the example code shown in Listing~\ref{lst:example}.
The example contains two synthetic vulnerabilities that are identical: program aborts while parsing (i) the input string literal {\tt doom}; (ii) the input whose cryptographic hash equals the string {\tt hash\_val}.
The {\tt abort} call following the cryptographic comparison is invoked via an alias called {\tt CUSTOM}.
We assume that the fuzzer is able to quickly find the crash due to the string literal comparison ({\tt doom}), but is unlikely to generate the input that satisfies the cryptographic operation.
This is a reasonable assumption since hash collisions are highly unlikely.
Faced with such a coverage bottleneck, we use the crash discovered by the fuzzer as a starting point for our vulnerability exploration, and proceed in three steps.
We first localize the fault underlying the observed crash by computing the set difference of program coverage traces for the crashing and non-crashing runs respectively.
Let us assume that the fuzzer corpus contains the crashing input {\tt doom} and it's parent mutation, say {\tt doo}.
Using basic block (line) coverage tracing which is fast to obtain, we compute the set difference of the faulty and non-faulty executions to be line $9$ (i.e., the {\tt abort} call).
Using the localized fault together with the crash stack trace, we then search for similar call sites using an automatically generated AST template.
Listing~\ref{lst:query} shows the template derived from the line of code containing a call to {\tt abort}, and two matches resulting from template matching.
Template matching results in two matches, one of which is the fuzzer-discovered vulnerability, and the other is a recurring instance {\it hiding} below cryptographic code.
Although textual pattern matching for vulnerable code patterns is possible, it breaks down when code properties are crucial to finding a match e.g., call to {\tt abort()} via an alias (line 5 of Listing~\ref{lst:example}).
Finally, we partially rank template matches by checking if the lines of code comprising a match have not been executed by a fuzzer (ranked high), or not (ranked low).
In our example, such a ranking would place the undiscovered bug on line $17$ of Listing~\ref{lst:example} above the bug on line $9$ that has been found by the fuzzer.

Leveraging static analysis to complement fuzzing is appealing for two reasons.
First, static analysis does not require a test harness, making it well-suited for our problem setting.
Second, by taking a program-centric view, static analysis provides a greater overall assurance on software quality or lack thereof.
Moreover, since we leverage concrete test cases to bootstrap our analysis, our vulnerability templates focus on a specific fault pattern that has occurred at least once with a demonstrable test input.
This begets greater confidence in the returned matches and a higher tolerance for false positives, from an analyst's point of view.

The proposed vulnerability exploration framework requires a coupling between dynamic and static analysis.
We begin by fuzzing a readily available program test case.
Subsequently, the following steps are taken to enable static exploration of fuzzer-determined program crashes.
\begin{itemize}
\item {\bf Fault localization}: We localize vulnerabilities (faults) reported by the fuzzer to a small portion of the code base using either a dynamic memory error detector such as AddressSanitizer~\cite{asanfull}, or using differential execution slices.
Fault localization serves as an interface for coupling dynamic and static analyses, and facilitates automatic generation of vulnerability templates.
\item {\bf Vulnerability Templates}: Using lines of code returned by the fault localization module, together with the crash stack trace, we automatically generate vulnerability templates.
The templates are encoded using code properties based on a program abstraction such as the abstract syntax tree (AST).
Template matching is used to for finding potentially recurring vulnerabilities.
\item {\bf Ranking Matches}: We rank matches returned by template matching before it is made available for human review.
Matches comprising lines of code not covered by fuzzing are ranked higher than those that have already been fuzzed.
\item {\bf Validation:} Finally, we manually audit the results returned by our analysis framework to ascertain if they can manifest as vulnerabilities in practice.
\end{itemize}

\subsection{Fault Localization}
\label{sec:code_suspects}


\begin{algorithm}[!t]
\caption{Pseudocode for execution slice based fault localization.}
\label{alg:slice_local}

\begin{algorithmic}[1]
\Function{obtain-slice}{$Input$, $Program$}
\LineComment Slice generated using coverage tracer
\State Return lines executed by $Program(Input)$
\EndFunction
\\
\Function{obtain-dice}{$Slice1$, $Slice2$}
\State dice = $Slice1$ - $Slice2$
\State \Return dice
\EndFunction
\\
\Function{localize-failure}{$Fault-Input$, $Program$, $Fuzz-Corpus$}
\State fault-slice = obtain-slice($Fault-Input$, $Program$)
\State nonfault-input = obtain-parent-mutation($Fault-Input$, $Fuzz-Corpus$)
\State nonfault-slice = obtain-slice($nonfault-input$, $Program$)
\State fault-dice = obtain-dice(fault-slice, nonfault-slice)
\State \Return fault-dice
\EndFunction
\end{algorithmic}
\end{algorithm}

Although a program stack trace indicates where a {\it crash} happened, it does not necessarily pin-point the root-cause of the failure.
This is because, a failure (e.g., memory access violation) manifests much after the trail of the faulty program instructions has been erased from the active program stack.
Therefore, fault localization is crucial for templating the root-cause of a vulnerability.

We localize a fuzzer-discovered program failure using a memory detector such as AddressSanitizer~\cite{asanfull}.
AddressSanitizer is a dynamic analysis tool that keeps track of the state of use of program memory at run time, flagging out-of-bounds reads/writes at the time of occurrence.
However, AddressSanitizer cannot localize failures {\it not} caused by memory access violations.
For this reason, we additionally employ a differential execution slicing~\footnote{An execution slice is the set of source lines of code/branches executed by a given input.} algorithm to localize general-purpose defects.

Agrawal et al.~\cite{agrawal1995fault} first proposed the use of differential execution slices (that the authors named execution dices) to localize a general-purpose program fault.
Algorithm~\ref{alg:slice_local} shows an overview of our implementation of this technique.
First, the execution slice for a faulty input is obtained ($fault-slice$, line 10 of Algorithm~\ref{alg:slice_local}).
Second, the fuzzer mutation that preceded the faulty input and did not lead to a failure is determined (line 11), and the execution slice for this input obtained (line 12).
Finally, the set difference of the faulty and the non-faulty execution slices is obtained (line 13).
This set difference is called the fault dice for the observed failure.
We obtain execution slices of a program using the SanitizerCoverage tool~\cite{coveragesan}.

In summary, fault localization helps us localize a fuzzer-discovered vulnerability to a small portion of the codebase.
Faulty code may then be used to automatically generate vulnerability templates.

\subsection{Vulnerability Templates}
\label{sec:traversals}


Faulty code snippets contain syntactic and semantic information pertaining to a program failure.
For example, the fact that dereference of the {\tt len} field from a pointer to {\tt struct udp} leads to an out-of-bounds memory access contains (i) the syntactic information that {\tt len} field dereference of a data-type {\tt struct udp} are potentially error-prone; and (ii) the semantic information that tainted input flows into the {\tt struct udp} type record, and that appropriate sanitization is missing in this particular instance.
Therefore, we leverage both syntactic, and semantic information to facilitate static exploration of fuzzer-determined program crashes.

Syntactic and semantic templates are derived from localized code snippets, and the crash stack trace.
Syntactic templates are matched against the program's abstract syntax tree (AST) representation, while semantic templates against the program's control flow graph (CFG) representation.
In the following, we briefly describe how templates are generated, and subsequently matched.

\paragraph{Syntactic Templates}

Syntactic templates are matched against the program abstract syntax tree (AST).
They may be formulated as functional predicates on properties of AST nodes.
We describe the process of formulating and matching AST templates using an out-of-bounds read in UDP parsing code of Open vSwitch v2.6.1 that was found by afl-fuzz and AddressSanitizer.

Listing~\ref{lst:ovs} shows the code snippet responsible for the out-of-bounds read.
The faulty read occurs on line 636 of Listing~\ref{lst:ovs} while dereferencing the {\tt udp\_header} struct field called {\tt udp\_len}.
The stack trace provided by AddressSanitizer is shown in Listing~\ref{lst:ovs_stack}.
In this instance, the fault is localized to the function named {\tt check\_l4\_udp}.
Post fault localization, a vulnerability (AST) template is derived from the AST of the localized code itself.

\begin{lstlisting}[style=minipage,firstnumber=624,xleftmargin=2em,label=lst:ovs,caption=Code snippet from Open vSwitch v2.6.1 that contains a buffer overread vulnerability in UDP packet parsing code.]
static inline bool
check_l4_udp(const struct conn_key *key, 
	     const void *data, size_t size,
	     const void *l3)
{
  const struct udp_header *udp = data;

(*@ \codehilig{-\hspace{1em}// Bounds check on data size missing} @*)
(*@ \codehilig{-\hspace{1em}if (size \textless{} UDP\_HEADER\_LEN) \{} @*)
(*@ \codehilig{-\hspace{2em}return false;} @*)
(*@ \codehilig{-\hspace{1em}\}} @*)

(*@ \codehilir{-\hspace{1em} size\_t udp\_len = ntohs(udp->udp\_len);} @*)

  if (OVS_UNLIKELY(udp_len < UDP_HEADER_LEN || 
		udp_len > size)) {
    return false;
  }
  ...
}
\end{lstlisting}

Listing~\ref{lst:ovs_ast} shows the AST fragment of the localized faulty code snippet, generated using the Clang compiler.
The AST fragment is a sub-tree rooted at the declaration statement on line 636, that assigns a variable named {\tt udp\_len} of type {\tt size\_t}, to the value obtained by dereferencing a struct field called {\tt udp\_len} of type {\tt const unsigned short} from a pointer named {\tt udp} that points to a variable of type to {\tt struct udp\_header}.
Using the filtered AST fragment, we use AST template matching to find similar declaration statements where {\tt udp\_len} is dereferenced.
The templates are generated by automatically parsing the AST fragment (as shown in Listing~\ref{lst:ovs_ast}), and creating Clang libASTMatcher~\cite{libastmatcher} style functional predicates.
Subsequently, template matching is done on the entire codebase.
Listing~\ref{lst:ovs_query} shows the generated template and the matches discovered.

\begin{lstlisting}[style=minipage,xleftmargin=0.5em,label=lst:ovs_stack,caption=Stack trace for the buffer overread in UDP packet parsing code obtained using AddressSanitizer.]
================================================
==48662==ERROR: AddressSanitizer:
heap-buffer-overflow on address 0x60600000ef3a at
 pc 0x0000005fd716 bp 0x7ffddc709c70 
 sp 0x7ffddc709c68

READ of size 2 at 0x60600000ef3a thread T0
    #0 0x5fd715 in check_l4_udp 
    lib/conntrack.c:636:33
    #1 0x5fcdcb in extract_l4 
    lib/conntrack.c:903:29
    #2 0x5f84bd in conn_key_extract
    lib/conntrack.c:978:13
    #3 0x5f78c4 in conntrack_execute
    lib/conntrack.c:304:14
    #4 0x56df58 in pcap_batch_execute_conntrack
    tests/test-conntrack.c:186:9
    ...
================================================
\end{lstlisting}

AST templates are superior to simple code searching tools such as {\tt grep} for multiple reasons.
First, they encode type information necessary to filter through only the relevant data types.
Second, they are flexible enough to mine for selective code fragments, such as searching for {\tt udp\_len} dereferences in binary operations in addition to declaration statements only.

\begin{lstlisting}[style=minipage,xleftmargin=1em,firstnumber=538,label=lst:ovs_match,caption=Match returned using automatically generated AST template shows a potentially recurring vulnerability in Open vSwitch 2.6.1. This new flaw was present in the portion of OvS code that lacked a test harness and was found during syntactic template matching.]
static void
pinctrl_handle_put_dhcpv6_opts(struct dp_packet
 *pkt_in, struct ofputil_packet_in *pin,
 struct ofpbuf *userdata, struct ofpbuf 
*continuation OVS_UNUSED)
{
...
    // Incoming packet parsed into udp struct
    struct udp_header *in_udp = 
			dp_packet_l4(pkt_in);
...
    // Dereference missing bounds checking
(*@ \codehilir{\hspace{2em}size\_t udp\_len = ntohs(in\_udp->udp\_len);} @*)

...
}
\end{lstlisting}

\begin{table*}[!tbh]
\begin{lstlisting}[style=minipage,xleftmargin=2em,label=lst:ovs_ast,caption=AST of the localized fault that triggers an out-of-bounds read in UDP packet parsing code. AST nodes of interest are shown in green.]
  |-DeclStmt 0x3232120 <line:636:5, col:41>
  | `-VarDecl lib/conntrack.c:636:12 used udp_len 'size_t':'unsigned long' cinit
...
...
  |             |         `-MemberExpr 0x32318f0 <lib/conntrack.c:636:28, col:33> 
(*@\codehilig{'const ovs\_be16':'const unsigned short' lvalue - \textgreater{} udp\_len} 0x3102ae0 @*)
  |             |           `-ImplicitCastExpr 0x32318d8 <col:28> 'const struct udp_header *' 
<LValueToRValue>
  |             |             `-DeclRefExpr 0x32318b0 <col:28> 'const struct udp_header *'
(*@lvalue Var 0x3231680 \codehilig{'udp' 'const struct udp\_header *'}@*)

\end{lstlisting}
\centering
\begin{lstlisting}[style=minipage,linewidth=15cm,xleftmargin=7em,label=lst:ovs_query,caption=AST template matching and its output. The code snippet surrounding match \#3 is shown in Listing~\ref{lst:ovs_match}.]
=============================== Query ===============================

let member memberExpr(allOf(hasDeclaration(namedDecl(hasName("udp_len"))), 
hasDescendant(declRefExpr(hasType(pointsTo(recordDecl(hasName("udp_header"))))))))

m declStmt(hasDescendant(member))

=============================== Matches ===============================

Match #1:

ovn/controller/pinctrl.c:635:5: note: "root" binds here
    out_ip6->ip6_ctlun.ip6_un1.ip6_un1_plen = out_udp->udp_len;
    ^~~~~~~~~~~~~~~~~~~~~~~~~~~~~~~~~~~~~~~~~~~~~~~~~~~~~~~~~~

Match #2:

tests/test-conntrack.c:52:9: note: "root" binds here
        udp->udp_src = htons(ntohs(udp->udp_src) + tid);
        ^~~~~~~~~~~~~~~~~~~~~~~~~~~~~~~~~~~~~~~~~~~~~~~

Match #3:

ovn/controller/pinctrl.c:550:5: note: "root" binds here
    size_t udp_len = ntohs(in_udp->udp_len);
    ^~~~~~~~~~~~~~~~~~~~~~~~~~~~~~~~~~~~~~~~

3 Matches.
\end{lstlisting}  
\end{table*}

Listing~\ref{lst:ovs_match} shows one of the matches discovered (see Match \#3 of Listing~\ref{lst:ovs_query}).
In the code snippet shown in Listing~\ref{lst:ovs_match}, the OVS controller function named {\tt pinctrl\_handle\_put\_dhcpv6\_opts} handles an incoming DHCP packet (containing a UDP packet) that is assigned to a pointer to {\tt struct udp\_header}, and subsequently dereferenced in the absence of a bounds-check on the length of the received packet.
This is one of the bugs found using syntactic template matching that was reported upstream, and subsequently patched by the vendor~\cite{ovsmatch}.
Moreover, this match alerted the OvS developers to a similar flaw in the DNS header parsing code.

To be precise, vulnerability templates need to encode both data and control flow relevant failure inducing code.
Otherwise, explicit sanitization of tainted input will be missed, leading to false positives.
To this end, we augment syntactic template matching with semantic (control and data-flow) template matching.

\paragraph{Semantic Templates}

Control and data-flow templates encode semantic code properties needed to examine the flow of tainted input.
However, since each defect is characterized by unique control and data-flow, semantic templates are harder to automate.
We remedy this problem by providing {\it fixed} semantic templates that are generic enough to be applied to any defect type.

We parse the program crash stack trace to perform semantic template matching.
First, we determine the function in which the program fails (top-most frame in the crash trace), and generate a template to match other call-sites of this function.
We call this a {\it callsite} template.
Callsite templates intuitively capture the insight that, if a program failure manifests in a given function, other calls to that function demand inspection.
Second, for memory access violation related vulnerabilities, we determine the data-type of the variable that led to an access violation, and assume that this data-type is {\it tainted}.
Subsequently, we perform taint analysis on this data-type terminating at pre-determined security-sensitive sinks such as {\tt memcpy}, {\tt strcpy} etc.
We call this a {\it taint} template.
Taint templates provide insight on risky usages of a data-type that is known to have caused a memory access violation.
Callsite and taint templates are matched against the program control flow graph (CFG).
They have been implemented as extensions to the Clang Static Analyzer framework~\cite{clangsa}.


\subsection{Match Ranking}
\label{sec:mine}


\begin{algorithm}[!t]
\caption{Pseudocode for ranking statically explored vulnerability matches.}
\label{alg:ranking}

\begin{algorithmic}[1]
\Function{isHigh}{$Matching-unit$, $Coverset$}
\State \For{each $m-unit$ in $Coverset$}
               \If{$m-unit$ == $Matching-unit$}
               \Return True
               \EndIf
\EndFor
\State \Return False
\EndFunction
\\
\Function{rank-matches}{$Matches$, $Coverset$}
\State RHigh = $\emptyset$
\State RLow = $\emptyset$
\State \For{each $match$ in $Matches$}
\If{isHigh($match$, $Coverset$)}
\State RHigh += $match$
\Else
\State RLow += $match$
\EndIf
\EndFor
\State \Return (RHigh, RLow)
\EndFunction
\end{algorithmic}
\end{algorithm}

Matches returned using static template matching may be used to (in)validate potentially recurring vulnerabilities in a codebase.
However, since vulnerability templates over-approximate failure-inducing code patterns, false positives are inevitable.
We remedy the false-positive problem using a simple yet practical match ranking algorithm.

Algorithm~\ref{alg:ranking} presents the pseudocode for our match ranking algorithm.
The procedure called {\tt RANK-MATCHES} accepts the set of template matches (denoted as $Matches$), and the set of program functions covered by fuzz testing (denoted as $Coverset$)  as input, and returns a partially orders list suitable for manual review.
For each match, we apply a ranking predicate on the program function in which the match was found.
We call this function, the {\it matching unit}.
The ranking predicate (denoted as the procedure $isHigh$) takes two input parameters: the matching function name, and the $Coverset$.
Under the hood, $isHigh$ simply performs a test of set membership; it checks if the matching unit is a member of the coverset, returning {\tt True} if it is a member, {\tt False} otherwise.
All matching units that satisfy the ranking predicate are ranked high, while the rest are ranked low.
The ranked list is returned as output.

Our ranking algorithm is implemented in Python using a hash table based data structure.
Therefore, ranking a match takes $O(1)$ on average, and $O(n)$ in the worst case, where $n$ is the number of functions in the coverset.
On average, the time to rank all matches grows linearly with the number of matches.
This is really fast in practice e.g., in the order of a few milliseconds (see Table~\ref{tab:runtime}).

Our prototype leverages {\it GCov}~\cite{gcov}, a publicly available program coverage tracing tool, for obtaining the coverset of a fuzzer corpus.
Although our prototype currently uses function as a matching unit, it may be suitably altered to work at the level of source line of code (basic blocks).
However, given that a fuzzer corpus typically contains thousands of test cases, we chose function-level tracing for performance reasons.

\subsection{Validation}
\label{sec:validate}


Although match ranking helps reduce the burden of false positives, it does not eliminate them entirely.
Therefore, we rely on manual audit to ascertain the validity of analysis reports.
Nonetheless, our approach focuses attention on recurrences of demonstrably vulnerable code patterns, thereby reducing the extent of manual code audit.

\section{Case Study: Open vSwitch}
\label{sec:results}
\begin{table*}[!tbh]
  \centering
  \begin{tabular}{p{6cm}lrr}
  \toprule
  Fuzzer-Discovered Vulnerability & CVE ID & Explored Matches & True Positives \\
  \toprule
 Out-of-bounds read (IP) & CVE-2016-10377~\cite{cve201610377} & 5 & 0 \\
 Out-of-bounds read (TCP) & CVE-2017-9264~\cite{cve20179264}  & 10 & 0 \\
 Out-of-bounds read (UDP) & CVE-2017-9264 & 2 & 1 \\
 Out-of-bounds read (IPv6)& CVE-2017-9264 & 3 & 0 \\
 Remote DoS due to assertion failure & CVE-2017-9214~\cite{cve20179214} & 22 & 0 \\
 Remote DoS due to unhandled packet & CVE-2017-9263~\cite{cve20179263} & 34 & 0 \\
 Out-of-bounds read & CVE-2017-9265~\cite{cve20179265} & 1 & 0 \\
  \bottomrule
  \multicolumn{2}{c}{Total} & 96 & 1 \\
  \bottomrule
  \end{tabular}
  \caption{Summary of static vulnerability exploration carried out on vulnerabilities found by fuzzing Open vSwitch. For each fuzzer-discovered vulnerability, our prototype generate a vulnerability template, and matched it against the entire codebase.}
  \label{tab:matches}
\end{table*}

\begin{table*}[!tbh]
  \centering
  \begin{tabular}{lrrr}
  \toprule
  CVE ID & Explored matches & Ranked high (untested) & Reduction in FP (in \%)\\
  \toprule
  CVE-2016-10377 & 5 & 0 & 100\\
  CVE-2017-9264  & 10 & 0 & 100 \\
  CVE-2017-9264 & 2 & 2 & 0 \\
  CVE-2017-9264 & 3 & 0 & 100 \\
  CVE-2017-9214 & 41 & 17 & 59 \\
  CVE-2017-9263 & 34 & 17 & 50 \\
  CVE-2017-9265 & 1 & 0 & 100 \\
  \bottomrule
  Total & 96 & 36 & 62 \\
  \bottomrule
  \end{tabular}
  \caption{Effectiveness of our matching ranking algorithm in highlighting untested code, and assisting in fast review of matches.}
  \label{tab:ranking}
\end{table*}

We evaluated our approach on multiple versions of Open vSwitch, an open-source virtual switch used in data centers.
We chose Open vSwitch for evaluation because (i) it is a good representative of production code; (ii) it has insufficient test harnesses suitable for fuzzing, resulting in program edge coverage of less than 5\%.

Our evaluations were performed using afl-fuzz for fuzzing, AddressSanitizer for fault localization, falling back to our implementation of differential slice-based fault localization, and our implementation of static template generation, matching, and ranking algorithms.
Experiments were carried out on a 64-bit machine with 80 CPU threads (Intel Xeon E7-4870) clocked at 2.4 GHz, and 512 GB RAM.

\paragraph{Fuzzing and Fault Localization}
Using the baseline fuzzer, we discovered multiple out-of-bounds reads and assertion failures in packet parsing code in Open vSwitch.
All the discovered flaws were triaged to ascertain their security impact, and subsequently reported upstream and fixed.
For each unique vulnerability, we used our fault localization module comprising AddressSanitizer, and differential execution slicing, to determine the lines of code triggering the vulnerability.

\paragraph{Template Matching}
Using localized code, we automatically generated a template suitable for matching similar code patterns elsewhere in the codebase.
For example, the AST snippet shown in Listing~\ref{lst:ovs_ast} was parsed to derive a template for CVE-2017-9264.
Subsequently, we used the tool {\tt clang-query} to perform template matching using the derived template.
Listing~\ref{lst:ovs_query} shows the outcome of template matching for one of the bugs comprising CVE-2017-9264.
For each vulnerability that the fuzzer discovered, we counted the number of matches (excluding the known vulnerability itself) returned using template matching.

We used semantic template matching only when syntactic template matching was too broad to capture the code pattern underlying the vulnerability.
For example, if a program crash was caused by a failed assertion, syntactic templates (that matched calls to all assertion statements), were augmented with semantic templates (that matched a smaller subset of assertion statements involving tainted data types).

\paragraph{Ranking}
The returned matches were ranked using our proposed ranking algorithm (see Algorithm~\ref{alg:ranking}), and the ranked output was used as a starting point for manual security audit.
Matches ranked high were reviewed first.
This enabled us to devote more time to audit untested code, than the code that had already undergone testing.

\begin{table*}[!tbh]
  \centering
  \begin{tabular}{lrrrrrp{2cm}}
  \toprule
  CVE ID & Localization & Syntactic & Semantic & Ranking & Total Run Time & Normalized \\
  \toprule
  CVE-2016-10377 & 82ms & 1.66s & -- & 63ms & 1.80s & 0.20x \\
  CVE-2017-9264 (TCP)  & 84ms & 3.20s & -- & 64ms & 3.34s & 0.25x \\
  CVE-2017-9264 (UDP) & 86ms & 4.77s & -- & 59ms & 4.91s & 0.37x\\
  CVE-2017-9264 (IPv6) & 91ms & 4.71s & -- & 60ms & 4.86s & 0.36x\\
  CVE-2017-9214 & 9ms & 8.44s & 44.17s & 60ms & 52.67s & 5.51x \\
  CVE-2017-9263 & 9ms & 11.88s & 44.26s & 59ms & 57.09s & 5.97x \\
  CVE-2017-9265 & 111ms & 5.74s & -- & 56ms & 5.9s & 0.62x\\
  \bottomrule
  \end{tabular}
  \caption{Run times of fault localization, template matching, and match ranking for all statically explored vulnerabilities in Open vSwitch. The absolute and relative (to code compilation) run times for our end-to-end analysis is presented in the final two columns. A normalized run time of $2$x denotes that our end-to-end analysis takes twice as long as code compilation.}
  \label{tab:runtime}
\end{table*}

\subsection{Analysis Effectiveness}

We evaluated the effectiveness of our approach in two ways: Quantifying (i) the raw false positive rate of our analysis; (ii) the benefit of the proposed ranking algorithm in reducing the effective false positive rate after match ranking was done.

To quantify the number of raw false positives, we counted the total number of statically explored matches, and the number of true positives among them.
A match was deemed a true positive if manual audit revealed that the tainted instruction underwent no prior sanitization and was thus potentially vulnerable.
Table~\ref{tab:matches} summarizes our findings.
Our prototype returned a total of 96 matches for the 7 vulnerabilities found by fuzzing (listed in column 1 of Table~\ref{tab:matches}).
Out of 96 matches, only one match corresponding to CVE-2017-9264 was deemed a new potential vulnerability.
This was reported upstream and subsequently patched~\cite{ovsmatch}.
Moreover, the reported (potential) vulnerability helped OvS developers uncover another similar flaw in the DHCPv6 parsing code that followed the patched UDP flaw.

Our ranking algorithm ranked untested code over tested code, thereby helping reduce the manual effort involved in validating potential false positives.
Although it is hard to correctly quantify the benefit of our ranking algorithm in bringing down the false positive rate, we employ a notion of {\it effective} false positive rate.
We define the effective false positive rate to be the false positive rate only among highly ranked matches.
This is intuitive, since auditing untested code is usually more interesting to a security analyst than auditing code that has already undergone testing.
Table~\ref{tab:ranking} summarizes the number of effective false positives due to our analysis.
In total, there were 36 matches (out of 96) that were ranked high, bringing down the raw false positive rate by 62\%.
Naturally, we confirmed that the single true positive was among the highly ranked matches.

Match ranking helps reduce, but not eliminate the number of false positives.
Indeed, 1 correct match out of 36 matches is very low.
Having said that, our approach has borne good results in practice, and has helped advance the tooling required for secure coding.
The additional patch that our approach contributed to is not the only way in which our approach met this objective.
We discovered that the template derived from the vulnerability CVE-2016-10377 present in an earlier version of Open vSwitch (v2.5.0), could have helped eliminate a similar vulnerability (CVE-2017-9264) in a later version (v2.6.1), perhaps during software development itself.
This shows that our approach is suitable for regression testing.
Indeed, OvS developers noted in personal communications with the authors that the matches returned by our tooling not only encouraged reasoning about corner cases in software development, but helped catch bugs (latent vulnerabilities) at an early stage.

\subsection{Analysis Runtime}

We quantified the run time of our tooling by measuring the total and constituent run times of our work-flow steps, starting from fault localization, and template matching, to match ranking.
Table~\ref{tab:runtime} presents our analysis run times for each of the fuzzer-discovered vulnerabilities in Open vSwitch.
Since fault localization was done using dynamic tooling (AddressSanitizer/coverage tracing), it was orders of magnitude faster (ranging between 9--111 milliseconds) than the time required for static template matching.
For each fuzzer-discovered vulnerability, we measured the template matching run time as the time required to construct and match the vulnerability template against the entire codebase.
Template matching run time comprised between 92--99\% of the end-to-end runtime of our tooling, and ranged from 1.8 seconds to 57.09 seconds.
Syntactic template matching was up to 4x faster than semantic template matching.
This conformed to our expectations, as semantic matching is slower due to the need to encode (and check) program data and control flow in addition to its syntactic properties.
Nonetheless, our end-to-end vulnerability analysis had a normalized run time (relative to code compilation time) of between 0.2x to 5.97x.
The potential vulnerability that our analysis pointed out in untested UDP parsing code, was returned in roughly a third of the time taken for code compilation of the codebase.
This shows that our syntactic analysis is fast enough to be applied on each build of a codebase, while our semantic analysis is more suitable to be invoked during daily builds.
Moreover, given the low run time of our analysis, templates derived from a vulnerability discovered in a given release may be continuously applied to future versions of the same codebase as part of regression testing.

\section{Related Work}
\label{sec:literature}
Our work brings together ideas from recurring vulnerability detection, and program analysis and testing.
In the following paragraphs, we compare our work to advances in these areas.

\paragraph{Patch-based Discovery of Recurring Vulnerabilities}
Redebug~\cite{jang2012redebug} and Securesync~\cite{pham2010detection} find recurring vulnerabilities by using syntax matching of templates derived from vulnerability patches.
Thus, {\it patched} vulnerabilities form the basis of their template-based matching algorithms.
In contrast, we template a vulnerability based on automatically localized failures, and debug information obtained from fuzzer reported crashes.
What makes our setting more challenging is the lack of a reliable code pattern (usually obtained from a patch) to build a template from.
As we have shown, it is possible to construct vulnerability templates even in this constrained environment and find additional vulnerabilities even {\it in the absence} of patches.

\paragraph{Code Clone Detection}
We are not the first to present a pattern-based approach to vulnerability detection.
Yamaguchi et al.~\cite{yamaguchi2012generalized} project vulnerable code patterns derived from patched vulnerabilities on to a vector space.
This permits them to extrapolate known vulnerabilities in current code, thereby permitting the discovery of recurring vulnerabilities.

Other researchers have focused on finding code clones regardless of them manifesting as vulnerabilities~\cite{bellon2007comparison, baxter1998clone,kontogiannis1996pattern,marcus2001identification}.
Code clone detection tools such as CPMiner~\cite{li2006cp}, CCFinder~\cite{kamiya2002ccfinder}, Deckard~\cite{jiang2007deckard} solve the problem of finding code clones but rely on sample code input to be provided.
These tools solve the more general problem of finding identical copies of user-provided code.
Although these tools serve as a building block for recurring vulnerability discovery, they require that the user specifies the code segment to be matched.
In a setting where a security analyst is auditing third-party code, manual specification of code templates might not be feasible.
By automatically performing template matching on fuzzer-discovered program crashes, leverage the fuzzer for the specification for vulnerable code patterns.

\paragraph{Hybrid Vulnerability Discovery}
SAGE~\cite{godefroid2012} is a white-box fuzz testing tool that combines fuzz testing with dynamic test-case generation.
Constraints accumulated during fuzz testing are solved using an SMT solver to generate test cases that the fuzzer alone could not generate.
This is expensive because it requires a sophisticated solver.
In a similar vein, Driller~\cite{stephens2016driller} augments fuzzing through selectively resorting to symbolic execution when fuzzer encounters coverage bottlenecks.
The use of symbolic execution to augment fuzzing is complementary to our approach.
In practice, security audits would benefit from both our approach as well as that proposed by prior researchers.

Saner~\cite{balzarotti2008} combines static and dynamic analyses towards identifying XSS and SQL injection vulnerabilities in web applications.
The authors of Saner use static analysis to capture a set of taint source-sink pairs from web application code, and subsequently use dynamic analysis on the captured pairs to tease out vulnerabilities.
Their evaluation on popular PHP applications show that dynamic analysis is able to bring down the number of false positives produced by static analysis, and find multiple vulnerabilities.
Like our work, Saner demonstrates that static and dynamic analyses can effectively complement each other.
In contrast to Saner, we differ in the order of analyses performed (we perform static analysis driven vulnerability exploration after confirmed taint source-sink pairs have been found), and in the target programming language.

Yamaguchi et al.~\cite{yamaguchi2015automatic} automatically infer search patterns for taint-style vulnerabilities from source code by combining static analysis and unsupervised machine learning.
They show that their approach helps reduce the amount of code audit necessary to spot recurring vulnerabilities by up to 94.9\%, enabling them to find 8 zero-day vulnerabilities in production software.
Their work is close in spirit to ours.
However, we avoid the computational overhead involved in their workflow (building a code property graph, pattern clustering etc.), while retaining their template matching run time.
In our framework, fault localization and result ranking run times are almost negligible.

\section{Conclusions and Future Work}
\label{sec:conclusion}
Fuzzing is a time-tested technique for discovering taint-style vulnerabilities in software.
However, fuzzing is mainly limited by test coverage, and the availability of fuzzable test cases.
In this paper, we leverage static analysis to perform an exhaustive search by using fuzzer-discovered vulnerabilities as a starting point.

We use fault localization techniques to narrow down the search for vulnerable code patterns.
Subsequently, localized code is used to automatically generate vulnerability templates.
False positives have been the primary drawback of static analysis tools.
As a remedy, we propose a ranking algorithm that brings attention to potential vulnerabilities in untested code.

We evaluate our approach on multiple versions of the Open vSwitch codebase, a popular virtual switch used in data centers.
Using static exploration of fuzzer-discovered vulnerabilities, we were able to discover an additional potential vulnerability in untested code.
Furthermore, we show that a vulnerability template derived from a dated vulnerability would have helped discover a recurring vulnerability in a later software release.
This shows that static vulnerability exploration has the potential to weed out flaws at an early stage of software development.
Indeed, our case study highlights the need to complement existing software testing approaches like fuzzing with static analysis.

Our work leaves open multiple avenues for future work.
At present, we rely on manual validation of statically discovered faults.
This may be complemented using selective symbolic execution tools such as {\it angr} so that additional diagnostics such as path reachability and concrete test input may be obtained.
Orthogonally, the precision of our templates can be improved by modeling data sanitization functions more precisely.

\bibliographystyle{abbrv}
\bibliography{master}





\end{document}